\pdfoutput=1

\documentclass[3p,times]{elsarticle}

\usepackage{ecrc}

\volume{00}

\firstpage{1}

\journalname{Nuclear Physics A}

\runauth{}

\jid{procs}

\jnltitlelogo{Nuclear Physics A}

\usepackage{amssymb}

\usepackage[figuresright]{rotating}

\newcommand{\Jpsi}{$J/\psi$ }
\newcommand{\pT}{$p_T$ }

\newcommand{\sNN}{$\sqrt{s_{_\mathrm{NN}}}$ }
\newcommand{\s}{$\sqrt{s}$ }
\newcommand{\pp}{$p$+$p$ }
\newcommand{\auau}{Au+Au }
\newcommand{\cucu}{Cu+Cu }
\newcommand{\raa}{$R_{AA}$ }

\begin{document}

\begin{frontmatter}



\dochead{}

\title{J/$\psi$ production at high $p_T$ at STAR}

\author[label1,label2]{Zebo Tang (for the STAR Collaboration)}
\address[label1]{Department of Modern Physics, University of Science and Technology of China, 96 Jinzhai Road, Hefei, Anhui, China 230026}
\address[label2]{Physics Department, Brookhaven National Laboratory, Upton, New York 11973, USA}



\begin{abstract}
We report results on $J/\psi$-hadron azimuthal angular
correlations in 200 GeV \pp collision in the STAR experiment at
RHIC. The extracted $B$-hadron feed-down contribution to inclusive
\Jpsi yield is found to be 10-25\% in $4<p_T<12 ~\textrm{GeV}/c$
and has no significant center-of-mass energy dependence from RHIC
to LHC. The \pT spectrum of charged hadron associated with
high-\pT \Jpsi triggers on the away side is found to be consistent
with that from di-hadron correlations. \Jpsi signal from partially
produced \auau 39 GeV data will also be presented to demonstrate
STAR's \Jpsi capability at RHIC low energy run.
\end{abstract}


\end{frontmatter}


\section{Introduction}
The dissociation of \Jpsi due to color-screening of their
constituent quarks in a Quark-Gluon Plasma (QGP) is a classic
signature for deconfinement in relativistic heavy-ion
collisions~\cite{colorscreen}. Results from the PHENIX experiment
at RHIC show that the suppression of \Jpsi as a function of
centrality (the number of participants) is similar to that
observed by NA50 and NA60 at the CERN-SPS, even though the
temperature and energy density reached in these collisions is
significantly lower than at RHIC~\cite{RHICSPS}. This indicates
that additional mechanisms, such as recombination of charm quarks
in the later stage of the collision and/or suppression of
feed-down contribution from charmonium excited states or
$B$-hadrons, may play an important role; they will need to be
studied systematically before conclusion from the observed
suppression pattern can be drawn. Recently, the STAR experiment
has extended \Jpsi suppression measurement to high \pT in \cucu
collisions and found that the \Jpsi nuclear modification factor
\raa is consistent with no \Jpsi suppression at
$p_T>5~\textrm{GeV}/c$, in contrast to the prediction from a
theoretical model of quarkonium dissociation in a strongly coupled
liquid using an AdS/CFT
approach~\cite{starHighPtJpsiPaper,adscft}. The project is not yet
complete and we need to increase the statistics, investigate the
meachanism of \Jpsi formation, and perform the same measurement
with a larger system (Au+Au). On the other hand, measurements from
CDF shows that the contribution of $B$-hadrons relative to the
inclusive \Jpsi yield in $p+\bar{p}$ collisions at 1.96 TeV
significantly increases with increasing $p_T$. The same
measurement at RHIC energy will be also essentially needed to
disentangle the physics origin of the high-\pT \Jpsi suppression
measurements~\cite{JpsiSpectra_CDFII}.

$B$ was rarely studied at RHIC in the past ten years. The
$B\rightarrow J/\psi$ measurements in heavy-ion collisions at STAR
are still difficult without a precise vertex detector. But it can
be done in \pp collisions through $J/\psi$-hadron correlations,
originally proposed and studied by UA1~\cite{UA1Simulation}.
Furthermore, $J/\psi$-hadron correlations can be also used to
study the hadronic activity produced in association with a
high-\pT \Jpsi to investigate its production mechanism which is
still poorly understood more than 30 years after the discovery of
$J/\psi$.

In this paper we present the measurement of the correlation
between high-\pT $J/\psi$'s and charged hadrons at mid-rapidity
with the STAR experiment in $p+p$ collisions at \s$=200
~\textrm{GeV}$ in RHIC year 2009 high luminosity run. We also
report the status of measurement of \Jpsi in \auau collisions at
\sNN$=39~\textrm{GeV}$ (an energy between CERN-SPS and RHIC top
energies) at STAR with newly fully-installed Time-Of-Flight (TOF)
detector~\cite{starTOF1,starTOF2,starTOF3}.

\section{high-$p_T$ \Jpsi production in \pp collisions at 200 GeV}
\begin{figure}[htb]
\begin{minipage}[t]{0.49\textwidth}
\includegraphics[height=0.98\textwidth, angle=90]{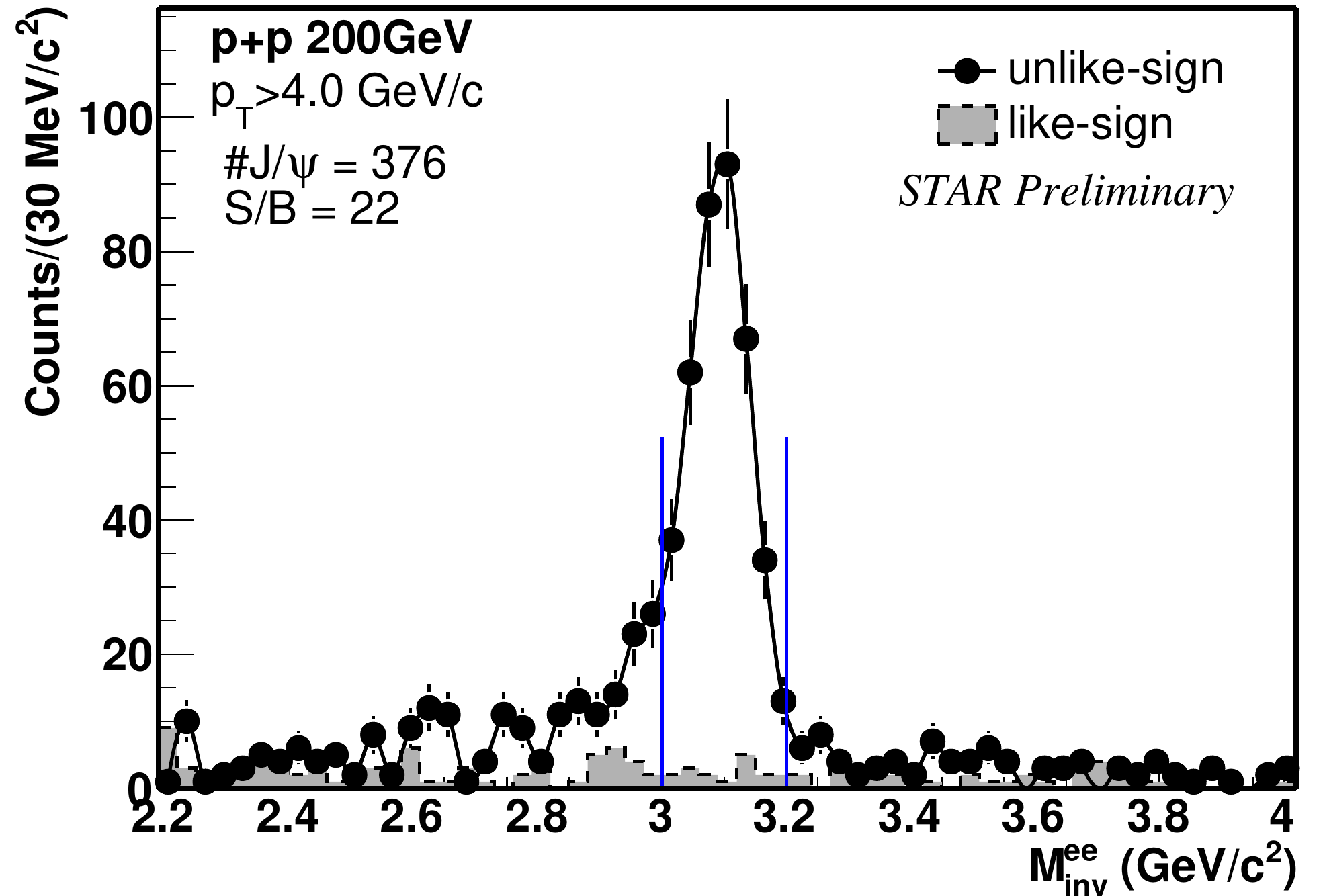}
\caption{Invariant mass distribution for unlike-sign (solid
circles) and like-sign (grey band) electron pairs at mid-rapidity
($|y|<1$) in \pp collisions at \sNN=200 GeV.} \label{fig:invmass}
\end{minipage}
\hspace{\fill}
\begin{minipage}[t]{0.49\textwidth}
\includegraphics[height=0.98\textwidth, angle=90]{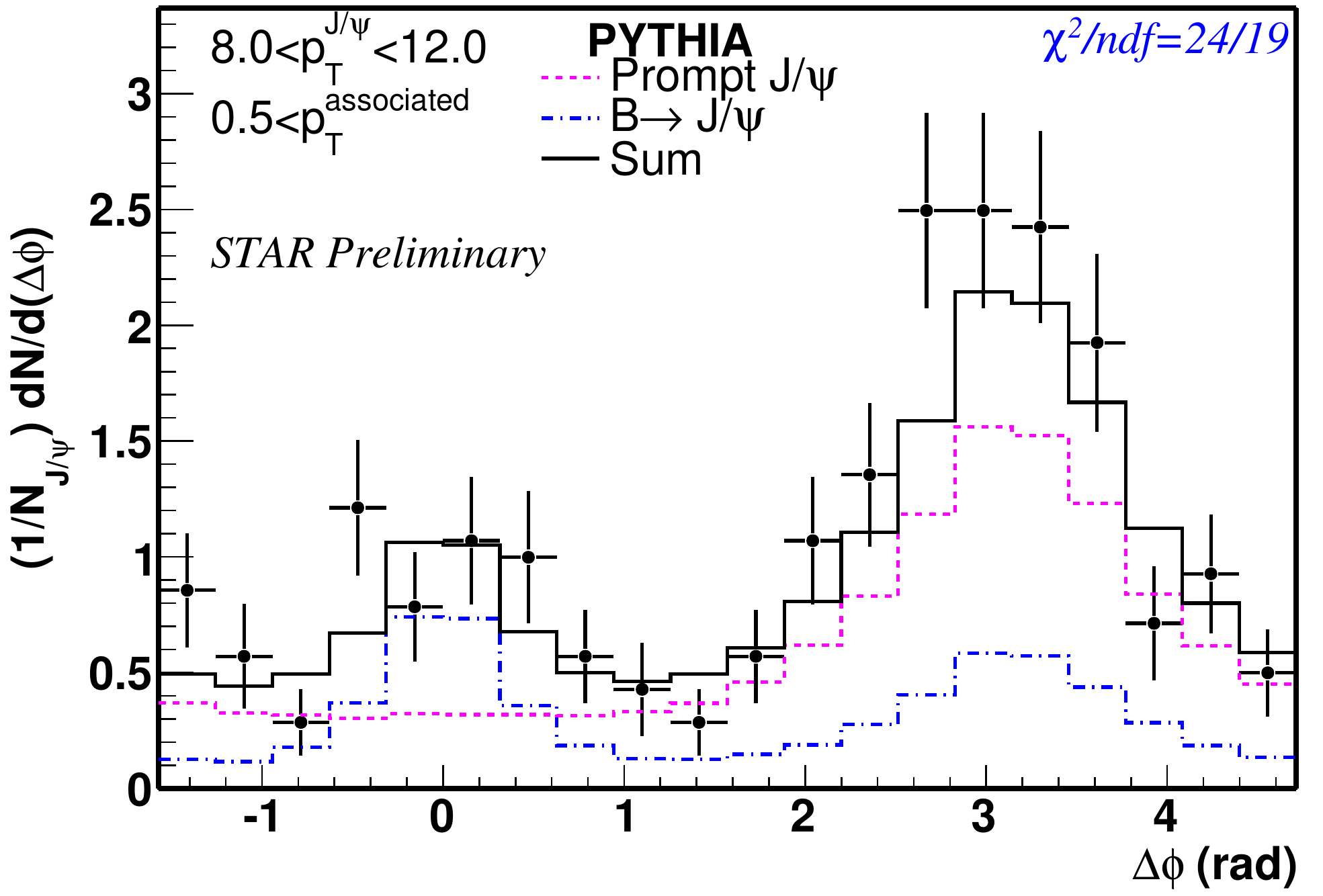}
\caption{$J/\psi$-hadron azimuthal angular correlations in the
\Jpsi \pT range of $8 <p_T<12 ~\textrm{GeV}/c$
 at mid-rapidity ($|y|<1$) in \pp collisions at \sNN=200 GeV.}\label{fig:corrFit}
\end{minipage}

\end{figure}

\begin{figure}[htb]
\begin{minipage}[t]{0.49\textwidth}\centering{
\includegraphics[height=0.9\textwidth, angle=90]{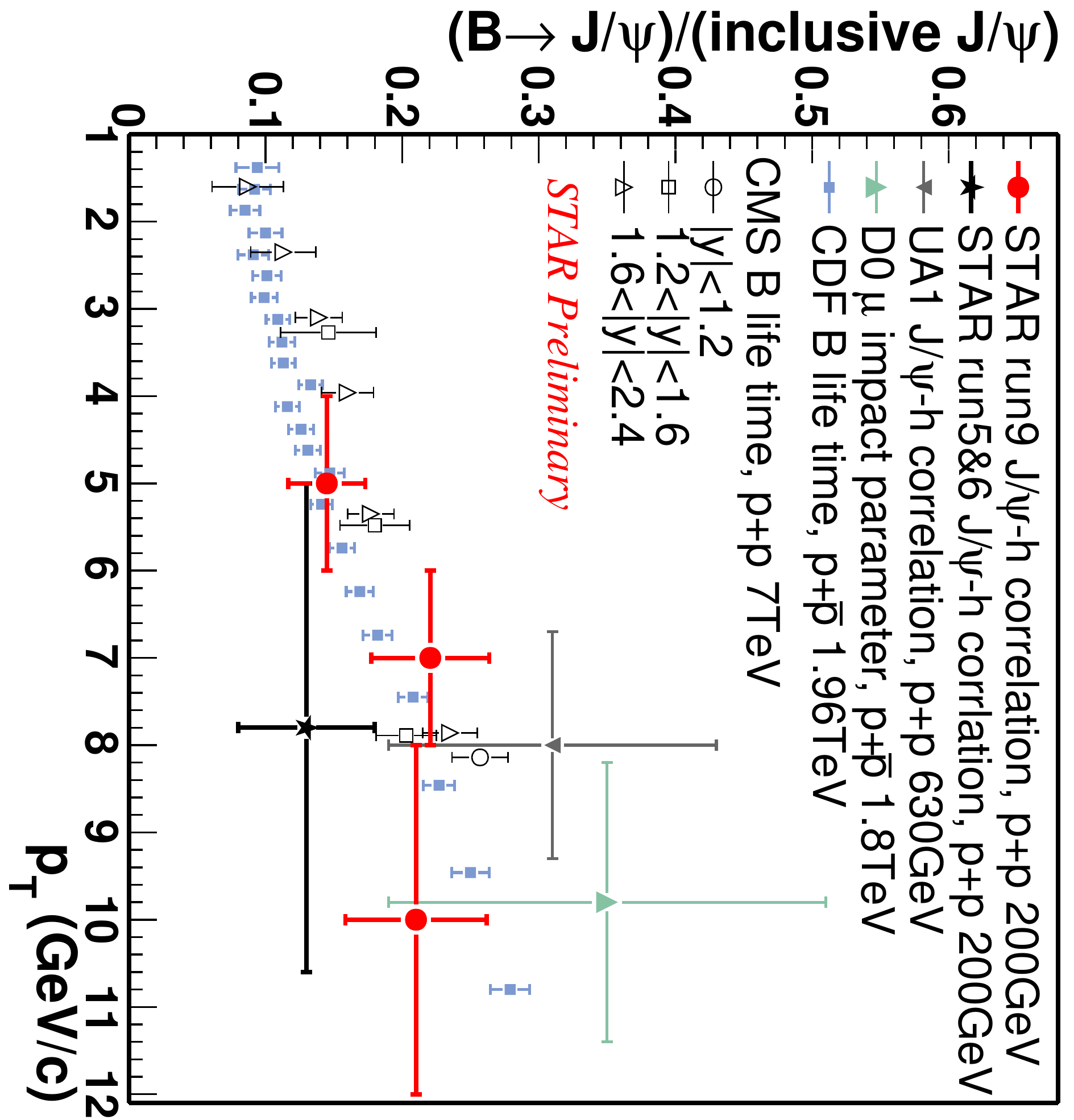}
\caption{Fraction of $B\rightarrow J/\psi$ over the inclusive
\Jpsi yield from two sets of run at STAR. The same ratios measured
by UA1, D0, CDF and CMS collaborations are also shown for
comparison.} \label{fig:B2Jpsi}}
\end{minipage}
\hspace{\fill}
\begin{minipage}[t]{0.49\textwidth}\centering{
\includegraphics[height=0.9\textwidth, angle=90]{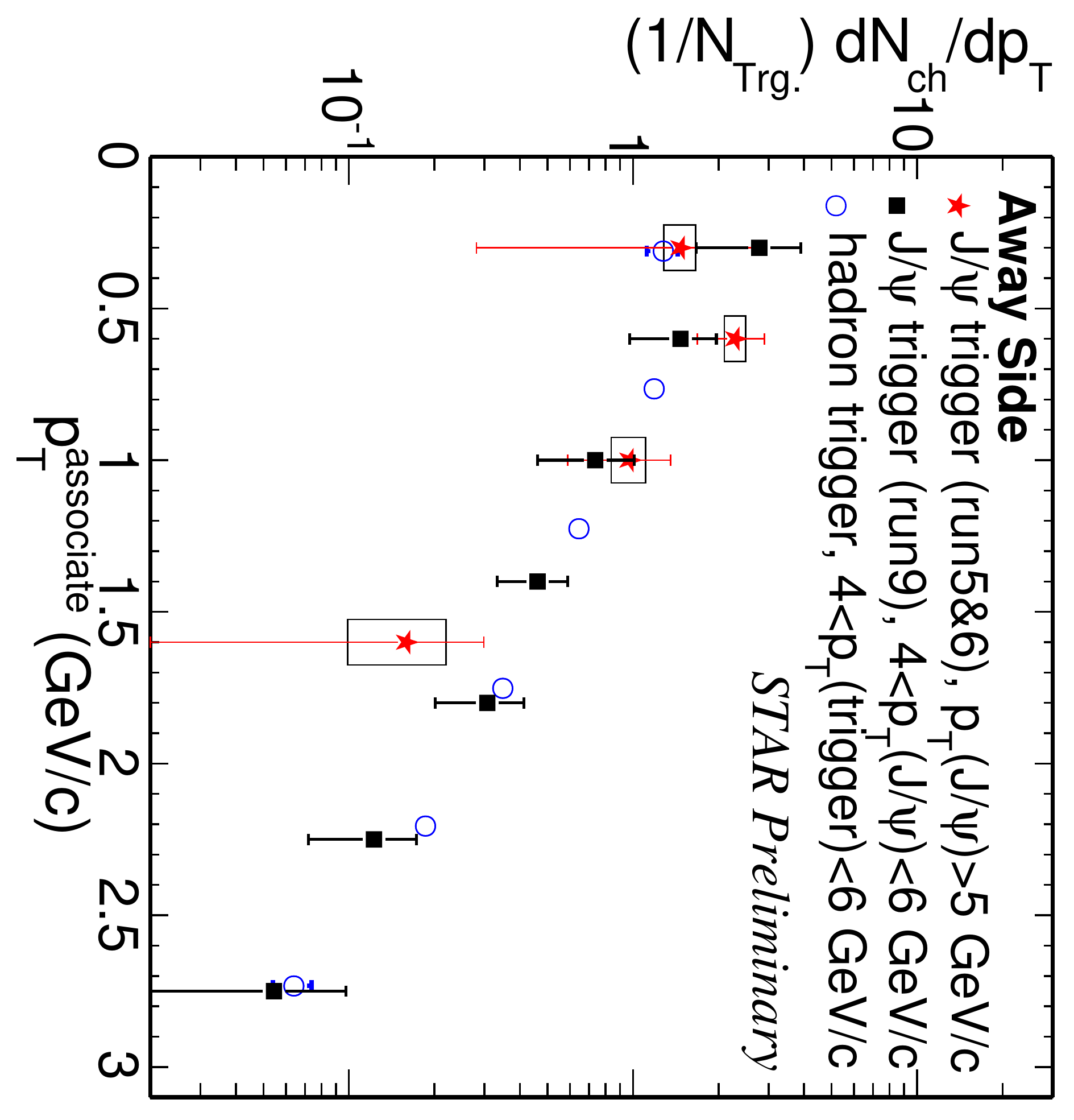}
\caption{Associated charged hadron \pT distributions on the away
side with respect to high-\pT \Jpsi triggers and charged hadron
triggers at mid-rapidity in \pp collisions at \sNN=200
GeV.}\label{fig:corrPt}}
\end{minipage}

\end{figure}

In this analysis, the \Jpsi is reconstructed through its decay
into electron-position pairs, $J/\psi \rightarrow e^+ e^-$
(Branching ratio (B) = 5.9\%). The data sample used was triggered
at level-0 by the STAR Barrel Electromagnetic Calorimeter (BEMC)
by requiring the transverse energy deposited in any tower ($\Delta
\eta \times \Delta \phi = 0.05 \times 0.05$) above a given
high-energy threshold to enrich high-\pT electrons. This
effectively enriches high-\pT $J/\psi$ with limited data
acquisition rate. The integrated luminosity is 1.8 $pb^{-1}$, 3.2
$pb^{-1}$ and 23.1 $pb^{-1}$ with transverse energy threshold $2.6
~\textrm{GeV}<E_T<4.3 ~\textrm{GeV}$, $E_T>4.3 ~\textrm{GeV}$ and
$E_T>6.0 ~\textrm{GeV}$ respectively. The reconstruction method is
similar as what we used in year 2005 and year 2006 data. We
tightened the $dE/dx$ cut slightly to enhance the
signal-to-background (S/B) ratio for the correlation
study~\cite{starHighPtJpsiPaper,zeboThesis}. In year 2009, STAR
installed 72\% TOF trays at mid-rapidity ($|\eta|<0.9$). This
detector combined with the Time Projection Chamber (TPC) can
clearly identify electrons from low to high $p_T$ by rejecting
hadrons at low and intermediate \pT range. To further improve the
S/B ratio of $J/\psi$, we also require the electron which does not
trigger the BEMC to have 1/$\beta$ measured by TOF within
0.97-1.03 when its \pT is less than 1
GeV/$c$~\cite{starTOFelectron}. Figure \ref{fig:invmass} shows the
invariant mass distribution for unlike-sign (solid circles) and
like-sign (shaded band) electron pairs. We reconstructed 376 \Jpsi
with $3.0 < M < 3.2 ~\textrm{GeV}/c^2$ at $p_T > 4
~\textrm{GeV}/c$. The S/B ratio in this range is 22. Such high S/B
ratio is very suitable for the $J/\psi$-hadron correlation study.
We do the correlation in 3 \Jpsi \pT slices: $4 - 6
~\textrm{GeV}/c$, $6 - 8 ~\textrm{GeV}/c$ and $8 - 12
~\textrm{GeV}/c$. Figure \ref{fig:corrFit} shows the azimuthal
angle correlations between high-\pT \Jpsi of $8 - 12
~\textrm{GeV}/c$ and charged hadrons. The correlated yield on the
near-side is not as significant at that in the di-hadron
correlation measurements~\cite{STAR_diHadron}. The lines show the
results of a PYTHIA calculation. The dot-dashed line exhibits a
strong near-side correlation compared to the away-side dominantly
from the decay $B \rightarrow J/\psi + X$. The solid line shows a
$\chi^2$ fit with the two simulated components to extract the
relative contribution of $B$-hadron feed-down to the inclusive
\Jpsi yield. This ratio is 10\%-25\% in the measured \pT range,
shown in Fig. \ref{fig:B2Jpsi} in red solid circles, increases
with increasing $p_T$. The results are consistent with STAR's
previous measurement (solid star symbol), but with better
precision~\cite{starHighPtJpsiPaper}. The same ratios measured by
UA1 in \pp collisions at 630 GeV, by D0 (CDF) in $p+\bar{p}$
collisions at 1.8 (1.96) TeV and by CMS in $p+p$ collisions at 7
TeV in various rapidity ranges are also shown for
comparison~\cite{JpsiSpectra_CDFII,UA1Simulation,D0Jpsi,CMSJpsi}.
They are consistent with each other even though the center-of-mass
energies differ by an order of magnitude. The ATLAS and LHCb
collaborations also observed a similar
behavior~\cite{LHCB2Jpsi,ATLASB2Jpsi}. The physics origin of this
consistency is still unclear. With such an amount of $B$-hadron
feed-down fraction, combined with this $J/\psi$-hadron correlation
study, further study of \Jpsi cross-section will allow us to
constrain the $B$ cross-section substantially in the future.

Figure \ref{fig:corrPt} shows the associated charged hadron \pT
distribution on the away side with respect to high-\pT \Jpsi
triggers and high-\pT charged hadron triggers. The \pT spectra of
charged hadron associated with high-\pT \Jpsi are consistent from
different runs, but year 2009 results have a better precision. To
compare the results with those from di-hadron correlation, we
require \Jpsi triggers in year 2009 run within the same \pT window
as charged hadron triggers: $4  - 6 ~\textrm{GeV}/c$. The \pT
spectra of the associated charged hadrons with respect to both
kinds of triggers are consistent with each other, which indicates
that the hadrons on the away side of \Jpsi triggers are dominantly
from light quark or gluon fragmentation, instead of heavy quark
fragmentation.

\section{\Jpsi production in \auau collisions at 39 GeV}
\begin{figure}[htb]
\begin{minipage}[t]{0.49\textwidth}
\includegraphics[height=0.98\textwidth, angle=90]{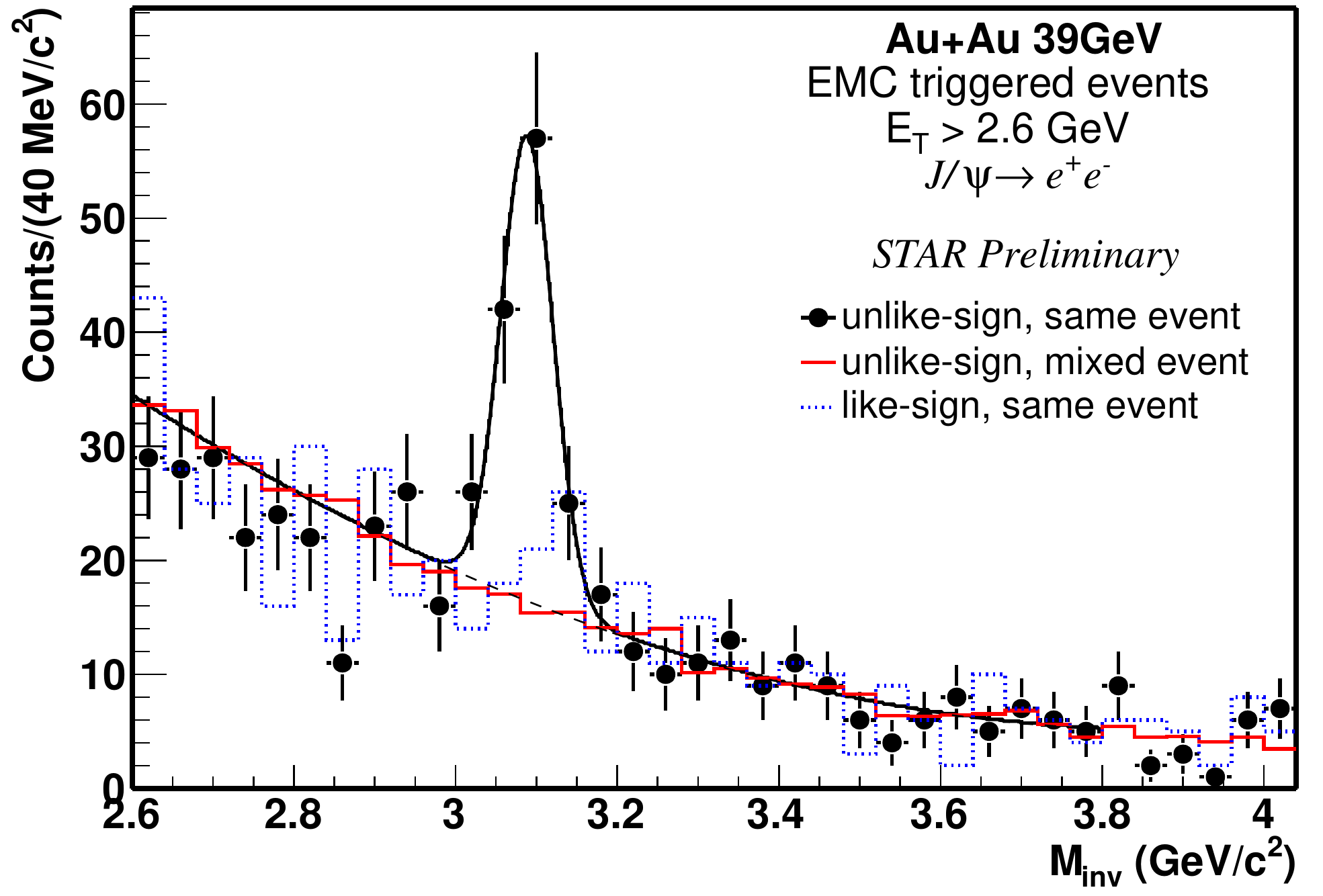}
\end{minipage}
\hspace{\fill}
\begin{minipage}[t]{0.49\textwidth}
\includegraphics[height=0.98\textwidth, angle=90]{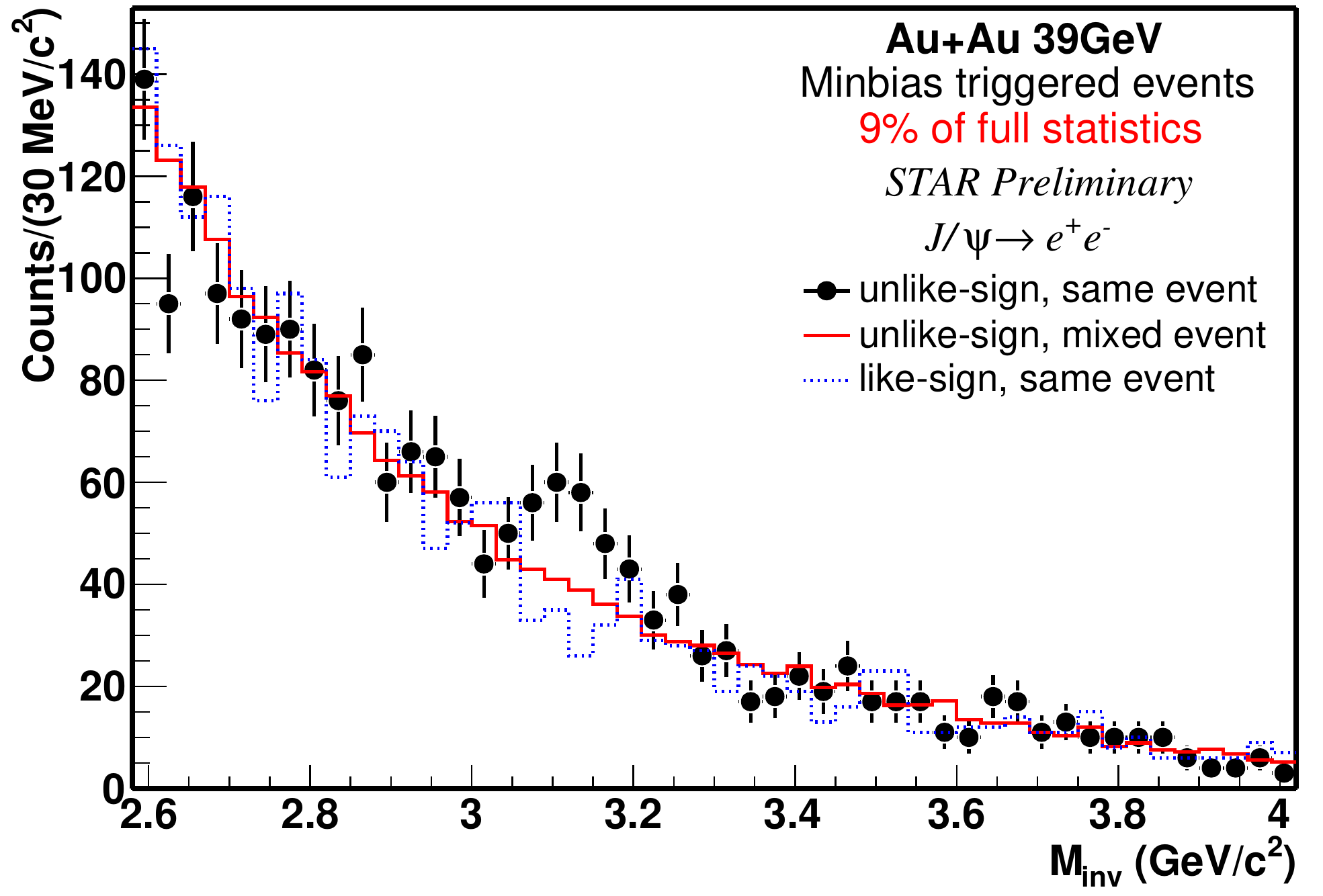}
\end{minipage}
\caption{Invariant mass distribution of electron pairs in BEMC
triggered (left) and minimum-bias (right) triggered \auau events
at \sNN=39 GeV. The solid and dashed histograms represent
background reproduced using like-sign and mixed-event technique
respectively.}\label{fig:invmassAuAu}
\end{figure}
The consistency of \Jpsi \raa at midrapidity at RHIC and SPS top
energies is still a puzzle. Two kinds of models with very
different physics origins (recombination models and sequential
dissociation models) can qualitatively explain this feature. The
measurements of \raa in heavy-ion collisions at a center-of-mass
energy between RHIC and SPS top energies are crucial to test these
models. The RHIC Beam Energy Scan (BES) program enables such
measurements (the reference data for \raa determination already
exist). STAR has recorded hundreds of million \auau events at \sNN
= 39, 62 and 200 GeV respectively during year 2010 run. Figure
\ref{fig:invmassAuAu} shows \Jpsi signal from partially produced
39 GeV \auau data to demonstrate STAR's \Jpsi capability at RHIC
low energy run.

The left panel of Fig. \ref{fig:invmassAuAu} shows the invariant
mass distributions for electron pairs in BEMC triggered events.
The electron identification and \Jpsi reconstruction is similar as
what we used in year 2009 \pp data. The S/B ratio is lower than
that in \pp collisions as expected, but still very high. To
improve the statistics, we also reproduce the combinatorial
background using mixed-event technique. It is consistent with that
from like-sign technique in the mass range shown in the figure. We
observed $82 \pm 13$ (6 $\sigma$) \Jpsi from this dataset, mainly
at $p_T>2$ GeV/$c$. To study \Jpsi production at low $p_T$, we
also analyzed minimum-bias (MB) triggered data. In this analysis,
we excluded BEMC from electron identification due to its
inefficiency at low $p_T$. The signal is shown in the right panel
of Fig. \ref{fig:invmassAuAu}. $91 \pm 22$ (4 $\sigma$) \Jpsi were
observed from this 9\% of full dataset, 52 in \pT range 0-2
GeV/$c$ and 39 in \pT range 2-4 GeV/$c$. We expect $\sim 1000$ (13
$\sigma$) \Jpsi signal from the full MB dataset. Our projection
shows STAR even has the capability to measure \Jpsi at 27 and 18
GeV with 1-2 weeks beam time in RHIC year 2011 run.

\section{Summary}
In summary, we reported results on $J/\psi$-hadron correlation in
\pp collisions at \s=200 GeV and \Jpsi signal in \auau collisions
at \sNN=39 GeV from the STAR experiment at RHIC. The fraction of
$B$-hadron feed-down contribution to inclusive \Jpsi yield in \pp
collisions was extracted from the $J/\psi$-hadron correlation and
found to be 10-25\% in $4<p_T<12 ~\textrm{GeV}/c$, with no
significant dependence on center-of-mass energy. The \pT spectra
of charged hadron associated with both high-\pT \Jpsi triggers and
high-\pT charged hadron triggers on the away side were found to be
consistent, which indicates the hadron production on the away side
is not dominantly from heavy quark fragmentation. STAR observed 6
$\sigma$ \Jpsi signal (mainly at $p_T>2 ~\textrm{GeV}/c$) in BEMC
triggered 39 GeV \auau events, and 4 $\sigma$ signal in 9\%
produced MB 39 GeV \auau events.

\section*{Acknowledgement}
The author is supported in part by the National Natural Science
Foundation of China under Grant No. 11005103 and the China
Fundamental Research Funds for the Central Universities.





\bibliographystyle{elsarticle-num}
\bibliography{HP2010_ZeboTang}








\end{document}